\documentclass{article}

\usepackage{PRIMEarxiv}

\usepackage[utf8]{inputenc} % allow utf-8 input
\usepackage[T1]{fontenc}    % use 8-bit T1 fonts
\usepackage{hyperref}       % hyperlinks
\usepackage{url}            % simple URL typesetting
\usepackage{booktabs}       % professional-quality tables
\usepackage{amsfonts}       % blackboard math symbols
\usepackage{nicefrac}       % compact symbols for 1/2, etc.
\usepackage{microtype}      % microtypography
\usepackage{lipsum}
\usepackage{fancyhdr}       % header
\usepackage{graphicx}       % graphics
\usepackage{amsmath}
\usepackage{amssymb}
\usepackage{amsthm}
\usepackage{verbatim}
\usepackage{footnote}
\usepackage{authblk}
\graphicspath{{media/}}     % organize your images and other figures under media/ folder

\title{BCRLB Under the Fusion Extended Kalman Filter
}

\author[1]{Mushen Lin}
\author[1]{Fenggang Yan}
\author[2]{Lingda Ren}
\author[1]{Xiangtian Meng $^{\ast,}$}
\author[3]{Maria Greco}
\author[3]{Fulvio Gini}
\author[1]{Ming Jin}
\affil[1]{School of Information Science and Engineering, Harbin Institute of Technology at Weihai, Werihai, China}
\affil[2]{Department of Electrical and Electronic Engineering, The University of Hong Kong, Hong Kong, China}
\affil[3]{Department of Information Engineering, University of Pisa, Pisa, Italy}

\date{} % 去掉日期

\begin{document}

\maketitle

\begin{abstract}
In the process of tracking multiple point targets in space using radar, since the targets are spatially well separated, the data between them will not be confused. Therefore, the multi-target tracking problem can be transformed into a single-target tracking problem. However, the data measured by radar nodes contains noise, clutter, and false targets, making it difficult for the fusion center to directly establish the association between radar measurements and real targets. To address this issue, the Probabilistic Data Association (PDA) algorithm is used to calculate the association probability between each radar measurement and the target, and the measurements are fused based on these probabilities. Finally, an extended Kalman filter (EKF) is used to predict the target states. Additionally, we derive the Bayesian Cramér-Rao Lower Bound (BCRLB) under the PDA fusion framework.
\end{abstract}

% keywords can be removed
\keywords{Bayesian Cramér-Rao lower bound \and Multi-target tracking \and Probabilistic data association}

% 1.系统模型
\section{System Model}
%\hspace{1.4em}
Consider a radar network with $M(M \geq 2)$ independent stationary radars, where the position of the $i$-th radar is denoted as $(x_i,y_i)$. 
Assume that there are \(Q\) targets within the surveillance region.
At tracking instant $k=0$, target $q$ is located at $(x_{0}^q,y_{0}^q)$ with the velocity of $(\Dot{x}_{0}^q,\Dot{y}_{0}^q)$.
It's worth emphasizing that the objective of the above radar networks is to select suitable beams and determine the optimal transmission power for the optimized tracking performance.  

% 1.1 状态转移模型
\subsection{State Transition Model}
%\hspace{1.4em}
The state vector of target \(q\) at instant \(k\) is represented as \(\mathbf{x}_k^q = [x_{Tk}^q, \dot{x}_{Tk}^q, y_{Tk}^q, \dot{y}_{Tk}^q]^\top \), where $T$ is the interval between two adjacent tracking frames. The state transition model at the \(k\)-th tracking frame is given by
\begin{equation}
    \mathbf{x}_k^q=\mathbf{F}_{q}\mathbf{x}_{k-1}^q+\mathbf{v}_k
\end{equation}
where $\mathbf{F}_{q}$ denotes the state transition function, $\mathbf{v}_k$ is the corresponding Gaussian process noise.

% 1.2 信号模型
\subsection{Signal Model}
The baseband echo received by the $i$-th radar node from the $q$-th target is \cite{15}
\begin{equation}
    r(t)=h_q\sqrt{\alpha_q P_{q,k}^i}s_i\left(t-\tau_q \right)e^{-j2\pi f_k^q\left(\mathbf{x}_k^q\right)t}+\omega(t)
\end{equation}
where $s_i(t)$ is the transmit signal of the radar $i$ and $\omega(t)$ is a zero-mean complex white Gaussian noise. $h_q$ denotes the reflectivity of the target $q$. $\alpha_q$ and $\tau_q$ represent the attenuation of the signal strength and the time delay, respectively. 

From the echoes, we extract the range, bearing angle, and Doppler frequency, which are described as follows \cite{16}:

% 1.3 量测模型
\subsection{Measurement Model}
Radar node $i$ is considered to receive an independent measurement $z_{q,k}^i = [r_{q,k}^i,f_{q,k}^i,\theta_{q,k}^i]$ for target $q$. The measurement model is given by \cite{14}
\begin{equation}
\mathbf{z}_{q,k}^i=\mathbf{G}_{q}\mathbf{x}_k^q+\mathbf{w}_{q,k}^i
\end{equation}
where $\mathrm{G}_q\left(\mathrm{x}_q^k\right)=\begin{bmatrix}r_{q,k}&\theta_{q,k}&f_{q,k}\end{bmatrix}^\top$ denotes the spatially consistent measurement function, which is described as follows
\begin{equation}
\left\{
\begin{array}{l}
r_{q,k}(\mathbf{x}_k^q)=\sqrt{\left(x_{k}^q\right)^2+\left(y_k^q\right)^2} \\
\theta_{q,k}(\mathbf{x}_k^q)=\arctan \left(\frac{y_k^q}{x_k^q}\right) \\
f_{q,k}(\mathbf{x}_k^q)=-\frac{2}{\lambda_{i}}\left({x}_{k}^{q}\cos\theta_{q,k}^i+{y}_{k}^{q}\sin\theta_{q,k}^i\right)
\end{array}\right.
\end{equation}
and $\mathbf{w}_{q,k}^i$ is the zero-mean, Gaussian measurement error with the covariance matrix $\mathbf R_{q,k}^i$ \cite{18} of radar $i$ with respect to target $q$ at the $k$-th frame
\begin{equation}
\mathbf R_{q,k}^i=blkdiag\left(\sigma^2_{r}(P_{q,k}^i),\sigma^2_{f}(P_{q,k}^i),\sigma^2_{\theta}(P_{q,k}^i)\right)
\end{equation}
where $\sigma_{r}(P_{q,k}^i)$, $\sigma_{f}(P_{q,k}^i)$ and $\sigma_{\theta}(P_{q,k}^i)$ are the estimation mean square error of target $q$'s range, Doppler frequency and bearing measurements, which are negatively correlated with each beam’s transmit power $P_{q,k}^i$.

% 2. 融合预测模型
\section{Fusion Prediction Model}
\label{sec:headings}
Considering that the targets are well separated spatially, the multi-target tracking problem is effectively reduced to a series of independent single-target tracking tasks. In this context, the probabilistic data association (PDA) algorithm is utilized to calculate the association probabilities for each measurement.
The fusion prediction process is primarily divided into the following three stages\cite{10}.

% 2.1 状态预测阶段
\subsection{State Prediction Stage}
State prediction stage calculates the prior probability density by using the previous posterior density
\begin{equation}
p\left(\mathrm{x}_k^q\mid\mathrm{z}_{1:k-1}^q\right)\sim\mathcal{N}\left(\mathrm{x}_{k|k-1}^q;\hat{\mathrm{x}}_{k|k-1}^q,\mathbf{\hat{P}}_{k|k-1}^q\right)
\end{equation}
where $\mathcal{N}\left(\mathrm{x}_{k|k-1}^q;\hat{\mathrm{x}}_{k|k-1}^q,\mathbf{\hat{P}}_{k|k-1}^q\right)$denotes the Gaussian distribution with mean $\hat{\mathrm{x}}_{k|k-1}^q$ and covariance $\mathbf{\hat{P}}_{k|k-1}^q$, whose mean and covariance can be written as
\begin{equation}
\hat{\mathrm{x}}_{k|k-1}^q=\mathrm{F}\hat{\mathrm{x}}_{k-1}^q,\quad\mathbf{\hat{P}}_{k|k-1}^q=\mathrm{Q}_T+\mathrm{F}\mathbf{\hat{P}}_{k-1}^q\mathrm{F}^\top
\end{equation}

% 2.2 数据关联阶段
\subsection{Data Association Stage}
The association probability $\beta_{q,k}^i$ is given by
\begin{equation}
\beta_{q,k}^i=\frac{P_{D,i}\cdot f_i(z_{q,k}^i|\hat{\mathrm{x}}_{k|k-1}^q)}{V_i\cdot \Lambda_i+P_{D,i}\cdot \sum_{i=1}^{M_q} f_i(z_{q,k}^i|\hat{\mathrm{x}}_{k|k-1}^q)}    
\end{equation}
where $M_q$ is the number of beams assigned to target, $V_i$ is the validation gate volume, $\Lambda_i$ is the clutter density, and $P_{D,i}$ is the probability of detection. The association probability is closely related to $f_i(z_{q,k}^i|\hat{\mathrm{x}}_{k|k-1}^q)$ , which represents the probability density function (PDF) of the measurement $z_{q,k}^i$ given the predicted state $\hat{\mathrm{x}}_{k|k-1}^q$ of target $q$.
\begin{equation}
f_i(z_{q,k}^i|\hat{\mathrm{x}}_{k|k-1}^q)=\frac{1}{\sqrt{(2\pi)^m|\mathrm{S}_{q,k}^i|}}\exp\left(-\frac{1}{2}\left(v_{q,k}^i\right)^\top\cdot \left(\mathrm{S}_{q,k}^i\right)^{-1}\left(v_{q,k}^i\right)\right)
\end{equation}
where $\nu_{q,k}^i=z_{q,k}^i-\mathbf{G}_{q}\left(\hat{\mathrm{x}}_{k|k-1}^q\right)$ is the innovation, and innovation covariance matrix $\mathrm{S}_{q,k}^i\left(P_{q,k}^i\right)$ is
\begin{equation}
\mathrm{S}_{q,k}^i\left(P_{q,k}^i\right)=\left(\nabla_{\hat{\mathbf{x}}_{k|k-1}^{q}}\mathbf{G}_{q}\right)\mathbf{\hat{P}}_{k|k-1}^q\left(\nabla_{\hat{\mathbf{x}}_{k|k-1}^{q}}\mathbf{G}_{q}\right)^\top+\mathbf{R}_{q,k}^i(P_{q,k}^i)
\end{equation}
The fused measurement $\bar{\mathbf{z}}_{q,k}=\sum_{i=1}^{M_q}\beta_{q,k}^iz_{q,k}^i$ and the fused measurement covariance matrix $\mathbb{R}_{q,k}$ can be written as
\begin{equation}
\mathbb{R}_{q,k}=\sum_{i=1}^{M_{q}}\beta_{q,k}^{i}\mathbf{R}_{q,k}^{i}+\sum_{i=1}^{M_{q}}\beta_{i,k}(z_{q,k}^{i}-\bar{\mathbf{z}}_{q,k})(z_{q,k}^{i}-\bar{\mathbf{z}}_{q,k})^{\top}
\end{equation}

% 2.3 状态更新阶段
\subsection{State Update Stage}
After obtaining the measurements, the state update stage is implemented to calculate the posterior density under the Bayes’ rule. Since $p\left(\mathrm{x}_k^q\mid\mathrm{z}_{1:k-1}^q\right)$ is Gaussian, it can be proved that $p\left(\mathrm{x}_{k}^q\mid\mathrm{z}_{q,1:k}\right)$ is also Gaussian and can be obtained by the Bayes’ rule:
\begin{equation}
p\left(\mathrm{x}_{k}^q\mid\mathrm{z}_{q,1:k}\right)\sim\mathcal{N}\left(\mathrm{x}_{k}^q;\hat{\mathrm{x}}_{k}^q,\mathbf{\hat{P}}_{k}^q\right)
\end{equation}
where
\begin{equation}
\begin{aligned}
\hat{\mathbf{x}}_{k}^{q}& \stackrel{\Delta}{=}\hat{\mathbf{x}}_{k}^{q}(P_{q,k}^{i}) \\
&=\hat{\mathrm{x}}_{k|k-1}^q+\mathrm{K}_k^q\left(P^i_{q,k}\right)\left[\bar{\mathbf{z}}_{q,k}-\mathbf{G}_{q}\left(\hat{\mathrm{x}}_{k|k-1}^q\right)\right]
\end{aligned}
\end{equation}
\begin{equation}
\begin{aligned}
\mathbf{\hat{P}}_k^q& \triangleq\mathbf{\hat{P}}_k^q(P_{q,k}^{i}) \\
&=\mathbf{\hat{P}}_{k|k-1}^q-\mathrm{K}_k^q\left(P^i_{q,k}\right)\mathbf{\hat{P}}_{k|k-1}^q\left(\nabla_{\hat{\mathbf{x}}_{k|k-1}^{q}}\mathbf{G}_{q}\right)
\end{aligned}
\end{equation}
in which $\mathrm{K}_k^q\left(P^i_{q,k}\right)=\mathbf{\hat{P}}_{k|k-1}^q\left(\nabla_{\hat{\mathbf{x}}_{k|k-1}^{q}}\mathbf{G}_{q}\right)\left[\mathrm{S}_k(P_{q,k}^i)\right]^{-1}$ is the Kalman gain, and covariance of the innovation $\mathrm{S}_k^q\left(P_{q,k}^i\right)$ given by
\begin{equation}
\mathrm{S}_k^q\left(P_{q,k}^i\right)=\left(\nabla_{\hat{\mathbf{x}}_{k|k-1}^{q}}\mathbf{G}_{q}\right)\mathbf{\hat{P}}_{k|k-1}^q\left(\nabla_{\hat{\mathbf{x}}_{k|k-1}^{q}}\mathbf{G}_{q}\right)^\top+\mathbb{R}_{q,k}
\end{equation}

% 3. BCRLB推导
\section{BCRLB Under the Fusion EKF}
In general, the mean-square-error (MSE) of any unbiased estimate is bounded by the BCRLB
\begin{equation}
\mathrm{E}\left\{\left[\mathrm{x}_k^q-\hat{\mathrm{x}}_k^q\left(P_{q,k}^i\right)\right]\left[\mathrm{x}_k^q-\hat{\mathrm{x}}_k^q\left(P_{q,k}^i\right)\right]^\top\right\}\geq\left[\mathbf{I}_{\mathbf{x}_k^q}\left(P_{q,k}^i\right)\right]^{-1}
\end{equation}
where $E \{\cdot\}$ is the mathematical expectation with respect to the joint density function of the states and measurements up to  time $k$. For the aforementioned Gaussian models, the FIM can be written as the sum of the following two components:
\begin{equation}
\mathbf{I}\left(\mathbf{x}_k^q,\mathbf{S}_k(q,:)\right)\approx\mathbf{I}_\mathrm{P}\left(\mathbf{x}_k^q\right)+\mathbf{I}_\mathrm{M}\left(\mathbf{x}_k^q,\mathbf{S}_k(q,:)\right)
\end{equation}
where $\mathbf{I}_\mathrm{P}\left(\mathbf{x}_k^q\right)$ corresponds to the prior information regarding the target state:
\begin{equation}
\mathbf{I}_\mathrm{P}\left(\mathbf{x}_k^q\right) = \left[\left(\nabla_{\mathbf{x}_k^q}\mathbf{F}_q\right)\left[\mathbf{I}\left(\mathbf{x}_{k-1}^q,\mathbf{S}_{k-1}(q,:)\right)\right]^{-1}\left(\nabla_{\mathbf{x}_k^q}\mathbf{F}_q\right)^\top+\mathbf{Q}_{k-1}^q\right]^{-1}
\end{equation}
and $\mathbf{I}_\mathrm{M}\left(\mathbf{x}_k^q,\mathbf{S}_k(q,:)\right)$ is the component arising from the measurements

\begin{align*}
    \mathbf{I}_\mathrm{M}\left({\mathbf{x}}_k^q,\mathbf{S}_k(q,:)\right)
    &=-\mathrm{E}\left[\frac{\partial^2\ln p(\bar{z}_{q,k}|\mathbf{x}_k^q)}{\partial\mathbf{x}_k^q(\partial\mathbf{x}_k^q)^\top}\right]
    \\&=\mathrm{E}\left\{\nabla_{{\mathbf{x}}_k^q}\mathbf{G}_q^\top\left[\mathbb{R}_{q,k}\left({\mathbf{x}}_k^q,\mathbf{S}_k(q,:)\right)\right]^{-1}\nabla_{{\mathbf{x}}_k^q}\mathbf{G}_q\right\}
\end{align*}

Since the measurement noise follows a Gaussian distribution, the conditional probability density function of the observation $\bar{z}_{q,k}$, given the state $\mathbf{x}_k^q$, is:
\begin{equation}
p(\bar{z}_{q,k}|\mathbf{x}_k^q)=\frac1{\sqrt{(2\pi)^m|\mathbb{R}_{q,k}|}}\exp\left(-\frac12[\bar{z}_{q,k}-\mathrm{G}_q(\mathbf{x}_k^q)]^\top\mathbb{R}_{q,k}^{-1}[\bar{z}_{q,k}-\mathrm{G}_q(\mathbf{x}_k^q)]\right)
\end{equation}
To simplify the process of calculating the expectation, which typically requires multiple Monte Carlo simulations, the statistical mean is often approximated by its state update value $\hat{\mathbf{x}}_{k}^{q}$.
\begin{equation}
\mathbf{I}_\mathrm{M}\left(\hat{\mathbf{x}}_k^q,\mathbf{S}_k(q,:)\right)\approx\left[\nabla_{\hat{\mathbf{x}}_k^q}\mathbf{G}_q^\top\left[\mathbb{R}_{q,k}\left(\hat{\mathbf{x}}_k^q,\mathbf{S}_k(q,:)\right)\right]^{-1}\nabla_{\hat{\mathbf{x}}_k^q}\mathbf{G}_q\right]
\end{equation}

\section{Conclusion}
In this paper, we present the prediction step using the fusion Extended Kalman Filter (EKF) based on the probabilistic data association (PDA) approach and provide proofs and other details for BCRLB Under the Fusion EKF.

%Bibliography
\bibliographystyle{unsrt}  
\bibliography{references}

\end{document}